\documentclass[apl,twocolumn,showpacs]{revtex4}
\usepackage{graphicx}
\usepackage{bm}
\usepackage{amsmath}
\usepackage{amssymb}
\newcommand{\etal}{{\it et al.}}
\renewcommand{\vec}[1]{{\bm{#1}}}

\begin{document}

\title{Enhancement of the Gilbert damping constant due to spin pumping
in non-collinear ferromagnet / non-magnet / ferromagnet trilayer systems}

\author{Tomohiro Taniguchi$^{1,2}$, Hiroshi Imamura$^{2}$}
\affiliation{${}^{1}$Institute for Materials Research, Tohoku University, Sendai 980-8577, \\
  $^{2}$Nanotechnology Research Institute, National Institute of Advanced Industrial Science and Technology, 
1-1-1 Umezono, Tsukuba, Ibaraki 305-8568, Japan}

\date{\today} 
\begin{abstract}
  {
    We analyzed the enhancement of the Gilbert damping constant due to
    spin pumping in non-collinear ferromagnet / non-magnet / ferromagnet
    trilayer systems. We show that the Gilbert damping constant depends
    both on the precession angle of the magnetization of the free layer
    and on the direction of the magntization of the fixed layer. We find
    the condition to be satisfied to realize strong enhancement of the
    Gilbert damping constant.
  }
\end{abstract}

\maketitle


There is currently great interest in the dynamics of magnetic
multilayers because of their potential applications 
in non-volatile magnetic random access memory (MRAM) and microwave devices. 
In the field of MRAM, 
much effort has been devoted to decreasing power consumption through 
the use of current-induced magnetization reversal (CIMR)
\cite{slonczewski96,berger96,kiselev03,deac05,kent04,lee05,seki06}.
Experimentally, CIMR is observed as 
the current perpendicular to plane-type giant magnetoresistivity (CPP-GMR) of a nano pillar,
in which the spin-polarized current injected from the fixed layer exerts 
a torque on the magnetization of the free layer. 
The torque induced
by the spin current is utilized to generate microwaves.


The dynamics of the magnetization $\vec{M}$ 
in a ferromagnet under an effective magnetic field $\vec{B}_{\mathrm{eff}}$ 
is described by the Landau-Lifshitz-Gilbert (LLG) equation
\begin{equation}
  \frac{d\vec{M}}{dt}=-\gamma\vec{M}\times\vec{B}_{\mathrm{eff}}+
  \alpha_{0}\frac{\vec{M}}{|\vec{M}|}\times\frac{d\vec{M}}{dt}\ ,
\end{equation}
where $\gamma$ and $\alpha_{0}$ are the gyromagnetic ratio and the Gilbert
damping constant intrinsic to the ferromagnet, respectively. The
Gilbert damping constant is an important parameter for spin
electronics since the critical current density of CIMR is proportional
to the Gilbert damping constant \cite{sun00,grollier03} and 
fast-switching time magnetization reversal is achieved for a large Gilbert damping constant \cite{koch04}. 
Several mechanisms intrinsic to ferromagnetic materials, 
such as phonon drag \cite{suhl98} and spin-orbit coupling \cite{Kambersky70}, 
have been proposed to
account for the origin of the Gilbert damping constant. 
In addition to these intrinsic mechanisms, 
Mizukami \etal \cite{mizukami02a,mizukami02b} 
and Tserkovnyak \etal \cite{tserkovnyak02a,tserkovnyak02b} 
showed that 
the Gilbert damping constant in a non-magnet (N) / ferromagnet (F) / non-magnet (N) trilayer system 
is enhanced due to spin pumping. 
Tserkovnyak \etal \cite{tserkovnyak03} also studied 
spin pumping in a collinear 
F/N/F trilayer system 
and showed that 
enhancement of the Gilbert damping constant depends on the precession angle of the magnetization of the free layer.


On the other hand, several groups 
who studied CIMR in a non-collinear F/N/F trilayer system 
in which the magnetization of the free layer is aligned 
to be perpendicular to that of the fixed layer
have reported the reduction of the critical current density \cite{kent04,lee05,seki06}. 
Therefore, it is intriguing to ask 
how the Gilbert damping constant is affected by spin pumping 
in non-collinear F/N/F trilayer systems.

\begin{figure}
\centerline{\includegraphics[width=0.7\columnwidth]{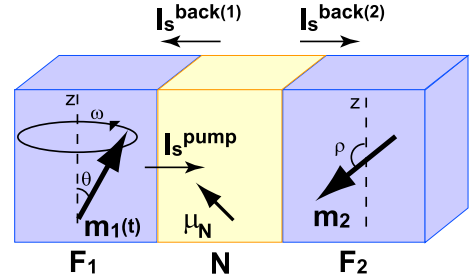}}
  \caption{(Color online) 
  The F/N/F trilayer system is schematically shown. 
  The magnetization of the ${\rm F}_{1}$ layer $(\vec{m}_{1})$ precesses
  around the $z$-axis with angle $\theta$ and angular velocity $\omega$.
  The magnetization of the ${\rm F}_{2}$ layer $(\vec{m}_{2})$ is
  fixed with tilted angle $\rho$. 
  The precession of the magnetization in the ${\rm F}_{1}$ layer pumps
  spin current $\vec{I}_{s}^{\mathrm{pump}}$ into the N and
  ${\rm F}_{2}$ layer, and creates the spin accumulation
  $\vec{\mu}_{\mathrm{N}}$ in the N layer.
  The spin accumulation induces the backflow spin current
  $\vec{I}_{s}^{\mathrm{back}(i)}(i=1,2)$. 
  }
\label{fig:fig1}
\end{figure}

In this paper, 
we analyze the enhancement of the Gilbert damping constant 
due to spin pumping in non-collinear F/N/F trilayer systems 
such as that shown in Fig. \ref{fig:fig1}. 
Following Refs. \cite{brataas01,tserkovnyak02a,tserkovnyak02b,tserkovnyak03}, 
we calculate the spin current 
induced by the precession of the magnetization of the free layer 
and the enhancement of the Gilbert damping constant. 
We show that 
the Gilbert damping constant depends 
not only on the precession angle $\theta$ of the magnetization of a free layer 
but also on the angle $\rho$ between the magnetizations of the fixed layer and the precession axis. 
The Gilbert damping constant is strongly enhanced 
if angles $\theta$ and $\rho$ satisfy the condition $\theta=\rho$ or $\theta=\pi-\rho$.

The system we consider is schematically shown in Fig. \ref{fig:fig1}.
A non-magnetic layer is sandwiched 
between two ferromagnetic layers, 
${\rm F}_{1}$ and ${\rm F}_{2}$. 
We introduce the unit vector $\vec{m}_{i}$ 
to represent the direction of the magnetization of the $i$-th ferromagnetic layer. 
The equilibrium direction of the magnetization $\vec{m}_{1}$ of the left free ferromagnetic layer ${\rm F}_{1}$ is
taken to exist along the $z$-axis. 
When an oscillating magnetic field is applied, 
the magnetization of the ${\rm F}_{1}$ layer precesses around the $z$-axis with angle $\theta$. 
The precession of the vector $\vec{m}_{1}$ is expressed as 
$\vec{m}_{1}=(\sin\theta\cos\omega t,\sin\theta\sin \omega t,\cos\theta)$, 
where $\omega$ is the angular velocity of the magnetization. 
The direction of the magnetization of the ${\rm F}_{2}$ layer, $\vec{m}_{2}$, 
is assumed to be fixed 
and the angle between $\vec{m}_{2}$ and the $z$-axis is represented by $\rho$.
The collinear alignment discussed in Ref. \cite{tserkovnyak03} corresponds to the case of $\rho=0,\pi$.


Before studying spin pumping in non-collinear systems, 
we shall give a brief review of the theory of spin pumping 
in a collinear F/N/F trilayer system \cite{tserkovnyak03}. 
Spin pumping is the inverse process of CIMR 
where the spin current induces the precession of the magnetization. 
Contrary to CIMR, 
spin pumping is the generation of the spin current induced by the precession of the magnetization. 
The spin current due to the precession of the magnetization in the ${\rm F}_{1}$ layer is given by
\begin{equation}
  \vec{I}_{s}^{\mathrm{pump}} 
  = \frac{\hbar}{4\pi}g^{\uparrow\downarrow}
  \vec{m}_{1}\times\frac{d\vec{m}_{1}}{dt}\ ,
\end{equation}
where 
$g^{\uparrow\downarrow}$ is a mixing conductance \cite{brataas00,brataas01} 
and $\hbar$ is the Dirac constant. 
Spins are pumped from the ${\rm F}_{1}$ layer into the N layer 
and the spin accumulation $\vec{\mu}_{\mathrm{N}}$ is created in the N layer.
Spins also accumulate in the ${\rm F}_{1}$ and ${\rm F}_{2}$ layers.
In the ferromagnetic layers 
the transverse component of the spin accumulation is assumed to be absorbed 
within the spin coherence length defined as 
$\lambda_{\mathrm{tra}} = \pi/|k_{\mathrm{F}_{i}}^{\uparrow} - k_{\mathrm{F}_{i}}^{\downarrow}|$, 
where
$k_{\mathrm{F}_{i}}^{\uparrow,\downarrow}$ is the spin-dependent Fermi wave number of the $i$-th ferromagnet. 
For ferromagnetic metals such as Fe, Co and Ni, 
the spin coherence length is a few angstroms \cite{stiles02}. 
Hence, the spin accumulation in the $i$-th ferromagnetic layer is aligned 
to be parallel to the magnetization,
i.e., 
$\vec{\mu}_{\mathrm{F}_{i}}=\mu_{\mathrm{F}_{i}}\vec{m}_{i}$.
The longitudinal component of the spin accumulation decays on the scale of spin diffusion length, 
$\lambda_{\mathrm{sd}}^{\mathrm{F}_{i}}$, 
which is of the order of 10 nm for typical ferromagnetic metals \cite{bass07}.


The difference in the spin accumulation of ferromagnetic and non-magnetic layers,
$\Delta\vec{\mu}_{i}=\vec{\mu}_{\mathrm{N}}-\mu_{\mathrm{F}_{i}}\vec{m}_{i}(i=1,2)$,
induces a backflow spin current, 
$\vec{I}_{s}^{\mathrm{back}(i)}$, 
flowing into both the ${\rm F}_{1}$ and ${\rm F}_{2}$ layers. 
The backflow spin current $\vec{I}_{s}^{\mathrm{back}(i)}$ is obtained using circuit theory \cite{brataas01} as
\begin{equation}
\begin{split}
    \vec{I}_{s}^{\mathrm{back}(i)}=
    \frac{1}{4\pi} 
    &
     \left\{ 
      \frac{2g^{\uparrow\uparrow}g^{\downarrow\downarrow}}
      {g^{\uparrow\uparrow}+g^{\downarrow\downarrow}}
      (\vec{m}_{i}\cdot\Delta\vec{\mu}_{i})\vec{m}_{i}
     \right. 
     \\
      & \left. +
      g^{\uparrow\downarrow}\vec{m}_{i}
      \times(\Delta\vec{\mu}_{i}\times\vec{m}_{i})\right\}\ , 
    \label{eq:backflow}
\end{split}
\end{equation}
where 
$g^{\uparrow\uparrow}$ and $g^{\downarrow\downarrow}$ are 
the spin-up and spin-down conductances, respectively. 
The total spin current flowing out of the ${\rm F}_{1}$ layer is given by
$\vec{I}_{s}^{\mathrm{exch}} = \vec{I}_{s}^{\mathrm{pump}} - \vec{I}_{s}^{\mathrm{back}(1)}$ \cite{tserkovnyak03}. 
The spin accumulation $\mu_{\mathrm{F}_{i}}$ in the ${\rm F}_{i}$ layer is
obtained by solving the diffusion equation. 
We assume that spin-flip scattering in the N layer is so weak 
that we can neglect the spatial variation of the spin current within the N layer,
$\vec{I}_{s}^{\rm exch}=\vec{I}_{s}^{{\rm back}(2)}$. 
The torque $\vec{\tau}_{1}$ acting on the magnetization of the ${\rm F}_{1}$ layer is given by 
$\vec{\tau}_{1} = \vec{I}_{s}^{\mathrm{exch}} - (\vec{m}_{1}\cdot\vec{I}_{s}^{\mathrm{exch}}) \vec{m}_{1} =
\vec{m}_{1}\times(\vec{I}_{s}^{\mathrm{exch}}\times\vec{m}_{1})$. 
For the collinear system, we have
\begin{equation}
  \vec{\tau}_{1}=
  \frac{g^{\uparrow\downarrow}}{8\pi}
  \left(
    1-\nu\frac{\sin^{2}\theta}{1-\nu^{2}\cos^{2}\theta}
  \right)
  \vec{m}_{1}\times\frac{d\vec{m}_{1}}{dt}\ ,
\end{equation}
where
$\nu=(g^{\uparrow\downarrow}-g^{*})/(g^{\uparrow\downarrow}+g^{*})$ is 
the dimensionless parameter introduced in Ref. \cite{tserkovnyak03}. 
The Gilbert damping constant in the LLG equation is 
enhanced due to the torque $\vec{\tau}_{1}$ as $\alpha_{0}\to\alpha_{0}+\alpha^{'}$ with
\begin{equation}
\alpha^{'}=
 \frac{g_{\mathrm{L}}\mu_{\mathrm{B}}g^{\uparrow\downarrow}}
 {8\pi M_{1} d_{\mathrm{F}_{1}}S}
 \left(1-\nu\frac{\sin^{2}\theta}{1-\nu^{2}\cos^{2}\theta}\right)\ ,
\label{eq:alpha1}
\end{equation}
where 
$g_{\mathrm{L}}$ is the Land\'e $g$-factor, 
$\mu_{\mathrm{B}}$ is the Bohr magneton, 
$d_{\mathrm{F}_{1}}$ is the thickness of the
${\rm F}_{1}$ layer and $S$ is the cross-section of the ${\rm F}_{1}$ layer.


Next, we move on to the non-collinear F/N/F trilayer system with $\rho=\pi/2$, 
in which the magnetization of the ${\rm F}_{2}$ layer is aligned to be perpendicular to the $z$-axis. 
Following a similar procedure, 
the LLG equation for the magnetization $\vec{M}_{1}$ in the ${\rm F}_{1}$ layer is expressed as
\begin{equation}
\frac{d\vec{M}_{1}}{dt}=
 -\gamma_{\mathrm{eff}}\vec{M}_{1}\times\vec{B}_{\mathrm{eff}}+
 \frac{\gamma_{\mathrm{eff}}}{\gamma}(\alpha_{0}+\alpha^{'})
 \frac{\vec{M}_{1}}{|\vec{M}_{1}|}\times\frac{d\vec{M}_{1}}{dt}\ ,
\label{eq:llg1}
\end{equation}
where 
$\gamma_{\mathrm{eff}}$ and $\alpha^{'}$ are 
the effective gyromagnetic ratio and 
the enhancement of the Gilbert damping constant, respectively. 
The effective gyromagnetic ratio is given by
\begin{equation}
  \gamma_{\mathrm{eff}}
  =
  \gamma
  \left(
    1- \frac{g_{\mathrm{L}}\mu_{\mathrm{B}}g^{\uparrow\downarrow}
      \nu\cot\theta\cos\psi\sin\omega t}
    {8\pi Md_{\mathrm{F}_{1}}S\epsilon}
  \right)^{-1}\ ,
  \label{eq:gamma1}
\end{equation}
where 
$\cos\psi=\sin\theta\cos\omega t=\vec{m}_{1}\cdot\vec{m}_{2}$ and 
\begin{equation}
  \epsilon
  = 1-\nu^{2}\cos^{2}\psi-
  \nu(\cot^{2}\theta\cos^{2}\psi-\sin^{2}\psi+\sin^{2}\omega t)\ .
  \label{eq:gamma2}
\end{equation}
The enhancement of the Gilbert damping constant is expressed as
\begin{equation}
  \alpha^{'}=
  \frac{g_{\mathrm{L}}\mu_{\mathrm{B}}g^{\uparrow\downarrow}}{8\pi Md_{\mathrm{F}_{1}}S}
  \left(1-\frac{\nu\cot^{2}\theta\cos^{2}\psi}{\epsilon}\right)\ . \label{eq:alpha2}
\end{equation}

\begin{figure}
  \centerline{\includegraphics[width=0.7\columnwidth]{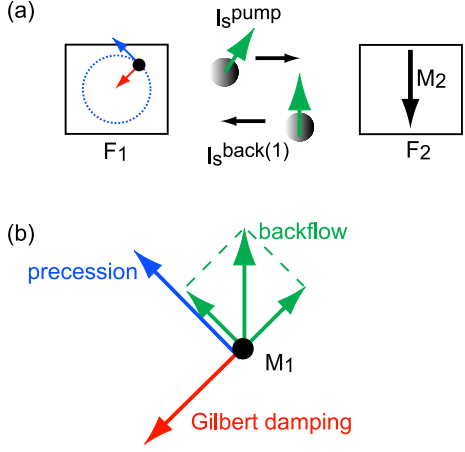}}
  \caption{(Color online) 
    (a) Top view of Fig. \protect\ref{fig:fig1}. 
    The dotted circle in ${\rm F}_{1}$ represents the precession of
    magnetization $\vec{M}_{1}$ and the arrow pointing to the center of
    this circle represents the torque of the Gilbert damping. The arrows in
    $\vec{I}_{s}^{\mathrm{pump}}$ and $\vec{I}_{s}^{\mathrm{back}(1)}$
    represent the magnetic moment of spin currents.
    (b) The back flow $\vec{I}_{s}^{\mathrm{back}(1)}$ has components
    aligned with the direction of the precession and the Gilbert damping.
  }
  \label{fig:fig2}
\end{figure}

It should be noted that, 
for non-collinear systems, 
both the gyromagnetic ratio and the Gilbert damping constant are modified by spin pumping, 
contrary to what occurs in collinear systems. 
The modification of the gyromagnetic ratio and the Gilbert damping constant due to spin pumping 
can be explained by considering the pumping spin current 
and the backflow spin current [See Figs. \ref{fig:fig2}(a) and \ref{fig:fig2}(b)]. 
The direction of the magnetic moment carried by the pumping spin current $\vec{I}_{s}^{\mathrm{pump}}$ is 
parallel to the torque of the Gilbert damping 
for both collinear and non-collinear systems. 
The Gilbert damping constant is enhanced 
by the pumping spin current $\vec{I}_{s}^{\mathrm{pump}}$. 
On the other hand, 
the direction of the magnetic moment 
carried by the backflow spin current 
$\vec{I}_{s}^{\mathrm{back}(1)}$ 
depends on the direction of the magnetization of the ${\rm F}_{2}$ layer. 
As shown in Eq. (\ref{eq:backflow}), 
the backflow spin current in the ${\rm F}_{2}$ layer $\vec{I}_{s}^{{\rm back}(2)}$ has a projection on $\vec{m}_{2}$.
Since we assume that 
the spin current is constant within the N layer, 
the backflow spin current in the ${\rm F}_{1}$ layer $\vec{I}_{s}^{\mathrm{back}(1)}$ 
also has a projection on $\vec{m}_{2}$. 
For the collinear system, 
both $\vec{I}_{s}^{\mathrm{pump}}$ and $\vec{I}_{s}^{\mathrm{back}(1)}$ are
perpendicular to the precession torque 
because $\vec{m}_{2}$ is parallel to the precession axis. 
However, 
for the non-collinear system, 
the vector $\vec{I}_{s}^{\mathrm{back}(1)}$ has a projection on the precession torque, 
as shown in Fig. \ref{fig:fig2}(b). 
Therefore, 
the angular momentum injected by $\vec{I}_{s}^{\mathrm{back}(1)}$ modifies 
the gyromagnetic ratio as well as the Gilbert damping in the non-collinear system.

Let us estimate the effective gyromagnetic ratio 
using realistic parameters. 
According to Ref. \cite{tserkovnyak03}, 
the conductances $g^{\uparrow\downarrow}$ and $g^{*}$ 
for a Py/Cu interface are given by $g^{\uparrow\downarrow}/S=15[\mathrm{nm}^{-2}]$ and $\nu\simeq 0.33$, respectively. 
The Land\'e $g$-factor is taken to be $g_{\mathrm{L}}=2.1$, 
magnetization is $4\pi M=8000$[Oe] 
and thickness $d_{\mathrm{F}_{1}}=5$[nm]. 
Substituting these parameters into Eqs. \eqref{eq:gamma1} and \eqref{eq:gamma2}, 
one can see that 
$ \left|\gamma_{\mathrm{eff}}/\gamma -1 \right| \simeq 0.001$. 
Therefore, 
the LLG equation can be rewritten as
\begin{equation}
  \frac{d\vec{M}_{1}}{dt}\simeq
  -\gamma\vec{M}_{1}\times\vec{B}_{\mathrm{eff}}
  +(\alpha_{0}+\alpha^{'})
  \frac{\vec{M}_{1}}{|\vec{M}_{1}|}\times\frac{d\vec{M}_{1}}{dt}.
  \label{eq:llg2}
\end{equation}
The estimated value of $\alpha^{'}$ is of the order of 0.001.
However, 
we cannot neglect $\alpha^{'}$ 
since it is of the same order 
as the intrinsic Gilbert damping constant $\alpha_{0}$  \cite{katine00,schreiber95}.

\begin{figure}
  \centerline{\includegraphics[width=0.7\columnwidth]{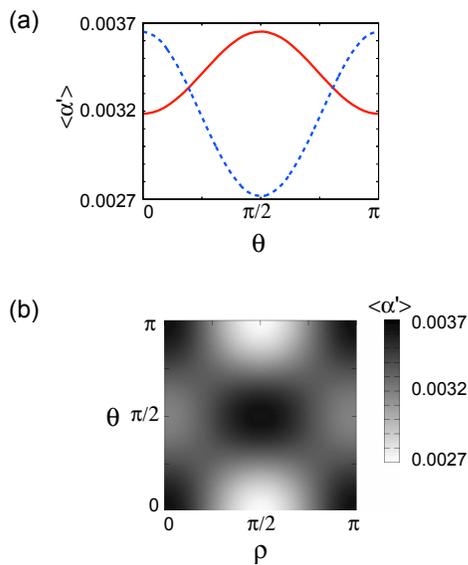}}
  \caption{(Color online) 
    (a) The time-averaged value of the enhancement of the Gilbert
    damping constant $\alpha^{'}$ is plotted as a function of the
    precession angle $\theta$.
    The solid line corresponds to the collinear system derived from
    Eq. \protect\eqref{eq:alpha2}. The dashed line corresponds to the
    non-collinear system derived from Eq. \protect\eqref{eq:alpha1}.
    (b) The time-averaged value of the enhancement of the Gilbert
    damping constant $\alpha^{'}$ of the non-collinear system 
    is plotted as a function of the precession angle $\theta$ and 
    the angle $\rho$ between the magnetizations of the fixed
    layer and the precession axis. 
  }
  \label{fig:fig3}
\end{figure}

Experimentally, 
the Gilbert damping constant is measured 
as the width of the ferromagnetic resonance (FMR) absorption spectrum. 
Let us assume that 
the ${\rm F}_{1}$ layer has no anisotropy 
and that an external field $\vec{B}_{\mathrm{ext}}=B_{0}\hat{\vec{z}}$ 
is applied along the $z$-axis. 
We also assume that 
the small-angle precession of the magnetization around the $z$-axis 
is excited by the oscillating magnetic field $\vec{B}_{1}$ applied in the $xy$-plane. 
The FMR absorption spectrum is obtained as follows \cite{vonsovskii}: 
\begin{equation}
  P=
  \frac{1}{T}\int_{0}^{T} dt 
  \frac{\alpha\gamma M\Omega^{2}B_{1}^{2}}
  {(\gamma B_{0}-\Omega)^{2}+(\alpha \gamma B_{0})^{2}}\ ,
\end{equation}
where 
$\Omega$ is the angular velocity of the oscillating magnetic field, 
$T=2\pi/\Omega$ and $\alpha=\alpha_{0}+\alpha^{'}$. 
Since $\alpha$ is very small, 
the absorption spectrum can be approximately expressed as 
$P \propto \alpha_{0} + \langle \alpha^{'} \rangle$ and
the highest point of the peak proportional to 
$\langle 1/(\alpha_{0} + \alpha^{'} )\rangle$, 
where $\langle \alpha' \rangle$ represents 
the time-averaged value of the enhancement of the Gilbert damping constant. 
In Fig. \ref{fig:fig3}(a), 
the time-averaged value 
$\langle \alpha^{'} \rangle$ 
for a non-collinear system 
in which $\rho=\pi/2$ is plotted by the solid line 
as a function of the precession angle $\theta$. 
The dotted line represents the enhancement of the Gilbert damping constant $\alpha^{'}$ 
for the collinear system given by Eq. \eqref{eq:alpha1}. 
The time-averaged value of the enhancement of the Gilbert damping constant 
$\langle \alpha' \rangle$ takes its maximum value 
at $\theta=0,\pi$ for the collinear system ($\rho=0,\pi$). 
Contrary to the collinear system, 
$\langle \alpha ' \rangle$ of the non-collinear system 
in which $\rho=\pi/2$ takes its maximum value at $\theta=\pi/2$.


As shown in Fig. \ref{fig:fig2}(b), 
the backflow spin current gives a negative contribution 
to the enhancement of the Gilbert damping constant. 
This contribution is given by 
the projection of the vector $\vec{I}_{s}^{\mathrm{back}(1)}$ 
onto the direction of the torque of the Gilbert damping, 
which is represented by the vector $\vec{m}_{1}\times\dot{\vec{m}}_{1}$. 
Therefore, 
the condition to realize the maximum value of the enhancement of the Gilbert damping is satisfied 
if the projection of $\vec{I}_{s}^{\mathrm{back}(1)}$ onto $\vec{m}_{1}\times\dot{\vec{m}}_{1}$
takes the minimum value; 
i.e., $\theta=\rho$ or $\theta=\pi-\rho$.

We can extend the above analysis to the non-collinear system 
with an arbitrary value of $\rho$. 
After performing the appropriate algebra, one can easily show
that the LLG equation for the magnetization of the ${\rm F}_{1}$ layer
is given by Eq. \eqref{eq:llg1} with
\begin{gather}
  \gamma_{\mathrm{eff}}
  =
  \gamma
  \left[1 -
    \frac{g_{\mathrm{L}}\mu_{\mathrm{B}} g^{\uparrow\downarrow} \nu \sin
      \rho \sin \omega t(\cot\theta\cos\tilde{\psi} -
      \csc\theta\cos\rho)}{8\pi MdS\tilde{\epsilon}}
  \right]^{-1} 
  \label{eq:gamma3}
  \\
  \alpha^{'}=\frac{g_{\mathrm{L}}\mu_{\mathrm{B}}g^{\uparrow\downarrow}}{8\pi MdS}
  \left\{
    1-\frac{\nu(\cot\theta\cos\tilde{\psi}-\csc\theta\cos\rho)^{2}}
    {\tilde{\epsilon}}\right\}\ ,
  \label{eq:alpha3}
\end{gather}
where $\cos\tilde{\psi} = \sin\theta\sin\rho\cos\omega t +
\cos\theta\cos\rho = \vec{m}_{1}\cdot\vec{m}_{2}$ and 
\begin{equation}
\begin{split}
    \tilde{\epsilon} = &
    1-\nu^{2}\cos^{2}\tilde{\psi} 
    \\
    &
    -\nu\{(\cot\theta\cos\tilde{\psi}-\csc\theta\cos\rho)^{2}
    -\sin^{2}\tilde{\psi}+\sin^{2}\rho\sin^{2}\omega
    t\}\ .
  \label{eq:gamma4}
\end{split}
\end{equation}
Substituting the realistic parameters into Eqs. \eqref{eq:gamma3} and \eqref{eq:gamma4}, 
we can show that 
the effective gyromagnetic ratio $\gamma_{\rm eff}$ 
can be replaced by $\gamma$ in Eq. \eqref{eq:llg1} 
and 
that the LLG equation reduces to Eq. \eqref{eq:llg2}. 
Figure \ref{fig:fig3}(b) shows 
the time-averaged value of 
the enhancement of the Gilbert damping constant $\langle \alpha^{'} \rangle$ of
Eq. \eqref{eq:alpha3}. 
Again, 
the Gilbert damping constant is strongly enhanced 
if angles $\theta$ and $\rho$ satisfy the condition 
that $\theta=\rho$ or $\theta=\pi-\rho$.


In summary, 
we have examined 
the effect of spin pumping on the dynamics of the magnetization of magnetic multilayers 
and calculated the enhancement of the Gilbert damping constant of 
non-collinear F/N/F trilayer systems due to spin pumping. 
The enhancement of the Gilbert damping constant depends 
not only on the precession angle $\theta$ of the magnetization of a free layer 
but also on the angle $\rho$ between the magnetizations of the fixed layer and the precession axis, 
as shown in Fig. \ref{fig:fig3}(b). 
We have shown that 
the $\theta$- and $\rho$-dependence of the enhancement of the Gilbert damping constant 
can be explained by analyzing the backflow spin current. 
The condition to be satisfied 
to realize strong enhancement of the Gilbert damping constant is $\theta=\rho$ or $\theta=\pi-\rho$.

The authors would like to acknowledge the valuable discussions we had with Y. Tserkovnyak,
S. Yakata, Y. Ando, S. Maekawa, S. Takahashi and J. Ieda. This work
was supported by CREST and by a NEDO Grant.



\begin{thebibliography}{22}
\expandafter\ifx\csname natexlab\endcsname\relax\def\natexlab#1{#1}\fi
\expandafter\ifx\csname bibnamefont\endcsname\relax
  \def\bibnamefont#1{#1}\fi
\expandafter\ifx\csname bibfnamefont\endcsname\relax
  \def\bibfnamefont#1{#1}\fi
\expandafter\ifx\csname citenamefont\endcsname\relax
  \def\citenamefont#1{#1}\fi
\expandafter\ifx\csname url\endcsname\relax
  \def\url#1{\texttt{#1}}\fi
\expandafter\ifx\csname urlprefix\endcsname\relax\def\urlprefix{URL }\fi
\providecommand{\bibinfo}[2]{#2}
\providecommand{\eprint}[2][]{\url{#2}}

\bibitem[{\citenamefont{Slonczewski}(1996)}]{slonczewski96}
\bibinfo{author}{\bibfnamefont{J.~C.} \bibnamefont{Slonczewski}},
  \bibinfo{journal}{J. Magn. Magn. Mater.} \textbf{\bibinfo{volume}{159}},
  \bibinfo{pages}{L1} (\bibinfo{year}{1996}).

\bibitem[{\citenamefont{Berger}(1996)}]{berger96}
\bibinfo{author}{\bibfnamefont{L.}~\bibnamefont{Berger}},
  \bibinfo{journal}{Phys. Rev. B} \textbf{\bibinfo{volume}{54}},
  \bibinfo{pages}{9353} (\bibinfo{year}{1996}).

\bibitem[{\citenamefont{Kiselev et~al.}(2003)\citenamefont{Kiselev, Sankey,
  Krivorotov, Emley, Schoelkopf, Buhrman, and Ralph}}]{kiselev03}
\bibinfo{author}{\bibfnamefont{S.~I.} \bibnamefont{Kiselev}},
  \bibinfo{author}{\bibfnamefont{J.~C.} \bibnamefont{Sankey}},
  \bibinfo{author}{\bibfnamefont{I.~N.} \bibnamefont{Krivorotov}},
  \bibinfo{author}{\bibfnamefont{N.~C.} \bibnamefont{Emley}},
  \bibinfo{author}{\bibfnamefont{R.~J.} \bibnamefont{Schoelkopf}},
  \bibinfo{author}{\bibfnamefont{R.~A.} \bibnamefont{Buhrman}},
  \bibnamefont{and} \bibinfo{author}{\bibfnamefont{D.~C.} \bibnamefont{Ralph}},
  \bibinfo{journal}{Nature} \textbf{\bibinfo{volume}{425}},
  \bibinfo{pages}{380} (\bibinfo{year}{2003}).

\bibitem[{\citenamefont{Deac et~al.}(2005)\citenamefont{Deac, Lee, Liu, Redon,
  Li, Wang, Nozi\'eres, and Dieny}}]{deac05}
\bibinfo{author}{\bibfnamefont{A.}~\bibnamefont{Deac}},
  \bibinfo{author}{\bibfnamefont{K.~J.} \bibnamefont{Lee}},
  \bibinfo{author}{\bibfnamefont{Y.}~\bibnamefont{Liu}},
  \bibinfo{author}{\bibfnamefont{O.}~\bibnamefont{Redon}},
  \bibinfo{author}{\bibfnamefont{M.}~\bibnamefont{Li}},
  \bibinfo{author}{\bibfnamefont{P.}~\bibnamefont{Wang}},
  \bibinfo{author}{\bibfnamefont{J.~P.} \bibnamefont{Nozi\'eres}},
  \bibnamefont{and} \bibinfo{author}{\bibfnamefont{B.}~\bibnamefont{Dieny}},
  \bibinfo{journal}{J. Magn. Magn. Mater.} \textbf{\bibinfo{volume}{290-291}},
  \bibinfo{pages}{42} (\bibinfo{year}{2005}).

\bibitem[{\citenamefont{Kent et~al.}(2004)\citenamefont{Kent, Ozyilmaz, and del
  Barco}}]{kent04}
\bibinfo{author}{\bibfnamefont{A.~D.} \bibnamefont{Kent}},
  \bibinfo{author}{\bibfnamefont{B.}~\bibnamefont{Ozyilmaz}}, \bibnamefont{and}
  \bibinfo{author}{\bibfnamefont{E.}~\bibnamefont{del Barco}},
  \bibinfo{journal}{Appl. Phys. Lett.} \textbf{\bibinfo{volume}{84}},
  \bibinfo{pages}{3897} (\bibinfo{year}{2004}).

\bibitem[{\citenamefont{Lee et~al.}(2005)\citenamefont{Lee, Redon, and
  Dieny}}]{lee05}
\bibinfo{author}{\bibfnamefont{K.~J.} \bibnamefont{Lee}},
  \bibinfo{author}{\bibfnamefont{O.}~\bibnamefont{Redon}}, \bibnamefont{and}
  \bibinfo{author}{\bibfnamefont{B.}~\bibnamefont{Dieny}},
  \bibinfo{journal}{Appl. Phys. Lett.} \textbf{\bibinfo{volume}{86}},
  \bibinfo{pages}{022505} (\bibinfo{year}{2005}).

\bibitem[{\citenamefont{Seki et~al.}(2005)\citenamefont{Seki, Mitani,
  Yakushiji, and Takanashi}}]{seki06}
\bibinfo{author}{\bibfnamefont{T.}~\bibnamefont{Seki}},
  \bibinfo{author}{\bibfnamefont{S.}~\bibnamefont{Mitani}},
  \bibinfo{author}{\bibfnamefont{K.}~\bibnamefont{Yakushiji}},
  \bibnamefont{and}
  \bibinfo{author}{\bibfnamefont{K.}~\bibnamefont{Takanashi}},
  \bibinfo{journal}{Appl. Phys. Lett.} \textbf{\bibinfo{volume}{89}},
  \bibinfo{pages}{172504} (\bibinfo{year}{2005}).

\bibitem[{\citenamefont{Sun}(2000)}]{sun00}
\bibinfo{author}{\bibfnamefont{J.~Z.} \bibnamefont{Sun}},
  \bibinfo{journal}{Phys. Rev. B} \textbf{\bibinfo{volume}{62}},
  \bibinfo{pages}{570} (\bibinfo{year}{2000}).

\bibitem[{\citenamefont{Grollier et~al.}(2003)\citenamefont{Grollier, Cros,
  Jaffres, Hamzic, George, Faini, Youssef, Gall, and Fert}}]{grollier03}
\bibinfo{author}{\bibfnamefont{J.}~\bibnamefont{Grollier}},
  \bibinfo{author}{\bibfnamefont{V.}~\bibnamefont{Cros}},
  \bibinfo{author}{\bibfnamefont{H.}~\bibnamefont{Jaffres}},
  \bibinfo{author}{\bibfnamefont{A.}~\bibnamefont{Hamzic}},
  \bibinfo{author}{\bibfnamefont{J.~M.} \bibnamefont{George}},
  \bibinfo{author}{\bibfnamefont{G.}~\bibnamefont{Faini}},
  \bibinfo{author}{\bibfnamefont{J.~B.} \bibnamefont{Youssef}},
  \bibinfo{author}{\bibfnamefont{H.~L.} \bibnamefont{LeGall}}, \bibnamefont{and}
  \bibinfo{author}{\bibfnamefont{A.}~\bibnamefont{Fert}},
  \bibinfo{journal}{Phys. Rev. B} \textbf{\bibinfo{volume}{67}},
  \bibinfo{pages}{174402} (\bibinfo{year}{2003}).

\bibitem[{\citenamefont{Koch et~al.}(2004)\citenamefont{Koch, Katine, and
  Sun}}]{koch04}
\bibinfo{author}{\bibfnamefont{R.~H.} \bibnamefont{Koch}},
  \bibinfo{author}{\bibfnamefont{J.~A.} \bibnamefont{Katine}},
  \bibnamefont{and} \bibinfo{author}{\bibfnamefont{J.~Z.} \bibnamefont{Sun}},
  \bibinfo{journal}{Phys. Rev. Lett.} \textbf{\bibinfo{volume}{92}},
  \bibinfo{pages}{088302} (\bibinfo{year}{2004}).

\bibitem[{\citenamefont{Suhl}(1998)}]{suhl98}
\bibinfo{author}{\bibfnamefont{H.}~\bibnamefont{Suhl}}, \bibinfo{journal}{IEEE
  Trans. Magn.} \textbf{\bibinfo{volume}{34}}, \bibinfo{pages}{1834}
  (\bibinfo{year}{1998}).

\bibitem[{\citenamefont{Kambersk{\'y}}(1970)}]{Kambersky70}
\bibinfo{author}{\bibfnamefont{V.}~\bibnamefont{Kambersk{\'y}}},
  \bibinfo{journal}{Can. J. Phys.} \textbf{\bibinfo{volume}{48}},
  \bibinfo{pages}{2906} (\bibinfo{year}{1970}).

\bibitem[{\citenamefont{Mizukami
  et~al.}(2002{\natexlab{a}})\citenamefont{Mizukami, Ando, and
  Miyazaki}}]{mizukami02a}
\bibinfo{author}{\bibfnamefont{S.}~\bibnamefont{Mizukami}},
  \bibinfo{author}{\bibfnamefont{Y.}~\bibnamefont{Ando}}, \bibnamefont{and}
  \bibinfo{author}{\bibfnamefont{T.}~\bibnamefont{Miyazaki}},
  \bibinfo{journal}{J. Magn. Magn. Mater.} \textbf{\bibinfo{volume}{239}},
  \bibinfo{pages}{42} (\bibinfo{year}{2002}{\natexlab{a}}).

\bibitem[{\citenamefont{Mizukami
  et~al.}(2002{\natexlab{b}})\citenamefont{Mizukami, Ando, and
  Miyazaki}}]{mizukami02b}
\bibinfo{author}{\bibfnamefont{S.}~\bibnamefont{Mizukami}},
  \bibinfo{author}{\bibfnamefont{Y.}~\bibnamefont{Ando}}, \bibnamefont{and}
  \bibinfo{author}{\bibfnamefont{T.}~\bibnamefont{Miyazaki}},
  \bibinfo{journal}{Phys. Rev. B} \textbf{\bibinfo{volume}{66}},
  \bibinfo{pages}{104413} (\bibinfo{year}{2002}{\natexlab{b}}).

\bibitem[{\citenamefont{Tserkovnyak and Brataas and Bauer}(2002)}]{tserkovnyak02a}
\bibinfo{author}{\bibfnamefont{Y.}~\bibnamefont{Tserkovnyak}} \bibnamefont{and}
  \bibinfo{author}{\bibfnamefont{A.}~\bibnamefont{Brataas}} \bibnamefont{and}
  \bibinfo{author}{\bibfnamefont{G.~E.~W.} \bibnamefont{Bauer}},
  \bibinfo{journal}{Phys. Rev. Lett.} \textbf{\bibinfo{volume}{88}},
  \bibinfo{pages}{117601} (\bibinfo{year}{2002}).

\bibitem[{\citenamefont{Tserkovnyak et~al.}(2002)\citenamefont{Tserkovnyak,
  Brataas, and Bauer}}]{tserkovnyak02b}
\bibinfo{author}{\bibfnamefont{Y.}~\bibnamefont{Tserkovnyak}},
  \bibinfo{author}{\bibfnamefont{A.}~\bibnamefont{Brataas}}, \bibnamefont{and}
  \bibinfo{author}{\bibfnamefont{G.~E.~W.} \bibnamefont{Bauer}},
  \bibinfo{journal}{Phys. Rev. B} \textbf{\bibinfo{volume}{66}},
  \bibinfo{pages}{224403} (\bibinfo{year}{2002}).

\bibitem[{\citenamefont{Tserkovnyak et~al.}(2003)\citenamefont{Tserkovnyak,
  Brataas, and Bauer}}]{tserkovnyak03}
\bibinfo{author}{\bibfnamefont{Y.}~\bibnamefont{Tserkovnyak}},
  \bibinfo{author}{\bibfnamefont{A.}~\bibnamefont{Brataas}}, \bibnamefont{and}
  \bibinfo{author}{\bibfnamefont{G.~E.~W.} \bibnamefont{Bauer}},
  \bibinfo{journal}{Phys. Rev. B} \textbf{\bibinfo{volume}{67}},
  \bibinfo{pages}{140404(R)} (\bibinfo{year}{2003}).

\bibitem[{\citenamefont{Brataas et~al.}(2001)\citenamefont{Brataas, Nazarov,
  and Bauer}}]{brataas01}
\bibinfo{author}{\bibfnamefont{A.}~\bibnamefont{Brataas}},
  \bibinfo{author}{\bibfnamefont{Y.~V.} \bibnamefont{Nazarov}},
  \bibnamefont{and} \bibinfo{author}{\bibfnamefont{G.~E.~W.}
  \bibnamefont{Bauer}}, \bibinfo{journal}{Eur. Phys. J. B}
  \textbf{\bibinfo{volume}{22}}, \bibinfo{pages}{99} (\bibinfo{year}{2001}).

\bibitem[{\citenamefont{Brataas et~al.}(2000)\citenamefont{Brataas, Nazarov,
  and Bauer}}]{brataas00}
\bibinfo{author}{\bibfnamefont{A.}~\bibnamefont{Brataas}},
  \bibinfo{author}{\bibfnamefont{Y.~V.} \bibnamefont{Nazarov}},
  \bibnamefont{and} \bibinfo{author}{\bibfnamefont{G.~E.~W.}
  \bibnamefont{Bauer}}, \bibinfo{journal}{Phys. Rev. Lett.}
  \textbf{\bibinfo{volume}{84}}, \bibinfo{pages}{2481} (\bibinfo{year}{2000}).

\bibitem[{\citenamefont{Stiles and Zangwill}(2002)}]{stiles02}
\bibinfo{author}{\bibfnamefont{M.~D.} \bibnamefont{Stiles}} \bibnamefont{and}
  \bibinfo{author}{\bibfnamefont{A.}~\bibnamefont{Zangwill}},
  \bibinfo{journal}{Phys. Rev. B} \textbf{\bibinfo{volume}{66}},
  \bibinfo{pages}{014407} (\bibinfo{year}{2002}).

\bibitem[{\citenamefont{Bass and Jr.}(2007)}]{bass07}
\bibinfo{author}{\bibfnamefont{J.}~\bibnamefont{Bass}} \bibnamefont{and}
  \bibinfo{author}{\bibfnamefont{W.~P.} \bibnamefont{Jr.}},
  \bibinfo{journal}{J. Phys.: Condens. Matter} \textbf{\bibinfo{volume}{19}},
  \bibinfo{pages}{183201} (\bibinfo{year}{2007}).

\bibitem[{\citenamefont{Katine et~al.}(2000)\citenamefont{Katine, Albert,
  Buhrman, Myers, and Ralph}}]{katine00}
\bibinfo{author}{\bibfnamefont{J.~A.}~\bibnamefont{Katine}},
  \bibinfo{author}{\bibfnamefont{F.~J.}~\bibnamefont{Albert}},
  \bibinfo{author}{\bibfnamefont{R.~A.}~\bibnamefont{Buhrman}},
  \bibinfo{author}{\bibfnamefont{E.~B.} \bibnamefont{Myers}}, \bibnamefont{and}
  \bibinfo{author}{\bibfnamefont{D.~C.}~\bibnamefont{Ralph}},
  \bibinfo{journal}{Phys. Rev. Lett.} \textbf{\bibinfo{volume}{84}},
  \bibinfo{pages}{3149} (\bibinfo{year}{2000}).

\bibitem[{\citenamefont{Schreiber et~al.}(1995)\citenamefont{Schreiber, Pflaum,
  Frait, M{\"u}hge, and Pelzl}}]{schreiber95}
\bibinfo{author}{\bibfnamefont{F.}~\bibnamefont{Schreiber}},
  \bibinfo{author}{\bibfnamefont{J.}~\bibnamefont{Pflaum}},
  \bibinfo{author}{\bibfnamefont{Th.} \bibnamefont{M{\"u}hge}}, \bibnamefont{and}
  \bibinfo{author}{\bibfnamefont{J.}~\bibnamefont{Pelzl}},
  \bibinfo{journal}{Solid State Commun.} \textbf{\bibinfo{volume}{93}},
  \bibinfo{pages}{965} (\bibinfo{year}{1995}).

\bibitem[{\citenamefont{Vonsovskii}(1964)}]{vonsovskii}
\bibinfo{editor}{\bibfnamefont{S.~V.} \bibnamefont{Vonsovskii}}, ed.,
  \emph{\bibinfo{title}{FERROMAGNETIC RESONANCE}} (\bibinfo{publisher}{Israel
  Program for Scientific Translations Ltd.}, \bibinfo{address}{Jersalem},
  \bibinfo{year}{1964}).

\end{thebibliography}
\end{document}